\documentclass[prl,nofootinbib,floatfix,twocolumn,showpacs,superscriptaddress]{revtex4}
\usepackage[dvips]{pstcol}
\usepackage{graphicx}
\usepackage{dcolumn}
\usepackage{bm}
\usepackage{rotating}
\usepackage{amsmath,amssymb,amsfonts}
\def\hermes{{\sc Hermes}}
\def\clas{{\sc Clas}}

\def\desy{{\sc Desy}}

\def\pythia6{{\sc Pythia6}}
\def\smh{\mbox{\large$\bm{\langle}$}\ensuremath{\sin(\phi\!-\!\phi_S)}
        \mbox{\large$\bm{\rangle}$}\ensuremath{_{\!U\!T}}^{\text{q}}}
\def\aulmh{\mbox{\large$\bm{\langle}$}\ensuremath{\sin\phi}
        \mbox{\large$\bm{\rangle}$}\ensuremath{_{\!U\!L}}^{\text{q}}}
\def\aulmpip{\mbox{\large$\bm{\langle}$}\ensuremath{\sin\phi}
        \mbox{\large$\bm{\rangle}$}\ensuremath{_{\!U\!L}}^{\text{q}, \pi^+}}
\def\aulmpim{\mbox{\large$\bm{\langle}$}\ensuremath{\sin\phi}
        \mbox{\large$\bm{\rangle}$}\ensuremath{_{\!U\!L}}^{\text{q}, \pi^-}}
\def\aulpmh{\mbox{\large$\bm{\langle}$}\ensuremath{\sin\phi}
        \mbox{\large$\bm{\rangle}$}\ensuremath{_{\!U\!L}}^{\text{l}}}
\def\cmph{\mbox{\large$\bm{\langle}$}\ensuremath{\sin(\phi\!+\!\phi_S)}
        \mbox{\large$\bm{\rangle}$}\ensuremath{_{\!U\!T}}^{\text{l}}}
\def\smph{\mbox{\large$\bm{\langle}$}\ensuremath{\sin(\phi\!-\!\phi_S)}
        \mbox{\large$\bm{\rangle}$}\ensuremath{_{\!U\!T}}^{\text{l}}}

\def\cmh{\mbox{\large$\bm{\langle}$}\ensuremath{\sin(\phi\!+\!\phi_S)}
        \mbox{\large$\bm{\rangle}$}\ensuremath{_{\!U\!T}}^{\text{q}}}
\newcommand{\pperp}{P_{h\perp}}
\newcommand{\Vpt}{\V{p_T^{}}}
\newcommand{\Vkt}{\V{k_T^{}}}
\newcommand{\scriptVpt}{\V{\scriptstyle p_T^{}}}
\newcommand{\scriptVkt}{\V{\scriptstyle k_T^{}}}
\newcommand{\Vpperp}{\V{\pperp^{}}}

\newcommand{\scriptVpperpu}{\V{\scriptstyle \hat{P}_{h\perp}^{}}}

\def\V#1{\mbox{\boldmath $#1$}}
\def\U#1{{\boldmath {\mathcal #1}}}

\begin{document}

\def\groupalberta{\affiliation{Department of Physics, University of Alberta, Edmonton, Alberta T6G 2J1, Canada}}
\def\groupargonne{\affiliation{Physics Division, Argonne National Laboratory, Argonne, Illinois 60439-4843, USA}}
\def\groupbari{\affiliation{Istituto Nazionale di Fisica Nucleare, Sezione di Bari, 70124 Bari, Italy}}
\def\groupbeijing{\affiliation{School of Physics, Peking University, Beijing 100871, China}}
\def\groupchina{\affiliation{Department of Modern Physics, University of Science and Technology of China, Hefei, Anhui 230026, China}}
\def\groupcolorado{\affiliation{Department of Physics, University of Colorado, Boulder, Colorado 80309-0390, USA}}
\def\groupdesy{\affiliation{DESY, 22603 Hamburg, Germany}}
\def\groupzeuthen{\affiliation{DESY, 15738 Zeuthen, Germany}}
\def\groupdubna{\affiliation{Joint Institute for Nuclear Research, 141980 Dubna, Russia}}
\def\grouperlangen{\affiliation{Physikalisches Institut, Universit\"at Erlangen-N\"urnberg, 91058 Erlangen, Germany}}
\def\groupferrara{\affiliation{Istituto Nazionale di Fisica Nucleare, Sezione di Ferrara and Dipartimento di Fisica, Universit\`a di Ferrara, 44100 Ferrara, Italy}}
\def\groupfrascati{\affiliation{Istituto Nazionale di Fisica Nucleare, Laboratori Nazionali di Frascati, 00044 Frascati, Italy}}
\def\groupgent{\affiliation{Department of Subatomic and Radiation Physics, University of Gent, 9000 Gent, Belgium}}
\def\groupgiessen{\affiliation{Physikalisches Institut, Universit\"at Gie{\ss}en, 35392 Gie{\ss}en, Germany}}
\def\groupglasgow{\affiliation{Department of Physics and Astronomy, University of Glasgow, Glasgow G12 8QQ, United Kingdom}}
\def\groupillinois{\affiliation{Department of Physics, University of Illinois, Urbana, Illinois 61801-3080, USA}}
\def\groupmit{\affiliation{Laboratory for Nuclear Science, Massachusetts Institute of Technology, Cambridge, Massachusetts 02139, USA}}
\def\groupmichigan{\affiliation{Randall Laboratory of Physics, University of Michigan, Ann Arbor, Michigan 48109-1040, USA }}
\def\groupmoscow{\affiliation{Lebedev Physical Institute, 117924 Moscow, Russia}}
\def\groupnikhef{\affiliation{Nationaal Instituut voor Kernfysica en Hoge-Energiefysica (NIKHEF), 1009 DB Amsterdam, The Netherlands}}
\def\groupstpetersburg{\affiliation{Petersburg Nuclear Physics Institute, St. Petersburg, Gatchina, 188350 Russia}}
\def\groupprotvino{\affiliation{Institute for High Energy Physics, Protvino, Moscow region, 142281 Russia}}
\def\groupregensburg{\affiliation{Institut f\"ur Theoretische Physik, Universit\"at Regensburg, 93040 Regensburg, Germany}}
\def\grouprome{\affiliation{Istituto Nazionale di Fisica Nucleare, Sezione Roma 1, Gruppo Sanit\`a and Physics Laboratory, Istituto Superiore di Sanit\`a, 00161 Roma, Italy}}
\def\groupsimonfraser{\affiliation{Department of Physics, Simon Fraser University, Burnaby, British Columbia V5A 1S6, Canada}}
\def\grouptriumf{\affiliation{TRIUMF, Vancouver, British Columbia V6T 2A3, Canada}}
\def\grouptokyo{\affiliation{Department of Physics, Tokyo Institute of Technology, Tokyo 152, Japan}}
\def\groupamsterdam{\affiliation{Department of Physics and Astronomy, Vrije Universiteit, 1081 HV Amsterdam, The Netherlands}}
\def\groupwarsaw{\affiliation{Andrzej Soltan Institute for Nuclear Studies, 00-689 Warsaw, Poland}}
\def\groupyerevan{\affiliation{Yerevan Physics Institute, 375036 Yerevan, Armenia}}


\groupalberta
\groupargonne
\groupbari
\groupbeijing
\groupchina
\groupcolorado
\groupdesy
\groupzeuthen
\groupdubna
\grouperlangen
\groupferrara
\groupfrascati
\groupgent
\groupgiessen
\groupglasgow
\groupillinois
\groupmit
\groupmichigan
\groupmoscow
\groupnikhef
\groupstpetersburg
\groupprotvino
\groupregensburg
\grouprome
\groupsimonfraser
\grouptriumf
\grouptokyo
\groupamsterdam
\groupwarsaw
\groupyerevan


\author{A.~Airapetian}  \groupmichigan
\author{N.~Akopov}  \groupyerevan
\author{Z.~Akopov}  \groupyerevan
\author{M.~Amarian}  \groupzeuthen \groupyerevan
\author{A.~Andrus}  \groupillinois
\author{E.C.~Aschenauer}  \groupzeuthen
\author{W.~Augustyniak}  \groupwarsaw
\author{R.~Avakian}  \groupyerevan
\author{A.~Avetissian}  \groupyerevan
\author{E.~Avetissian}  \groupfrascati
% ...

\collaboration{The HERMES Collaboration} \noaffiliation

\title{Subleading-twist effects in single-spin asymmetries in
semi-inclusive deep-inelastic scattering on a longitudinally polarized 
hydrogen target }

\date{\today}

\begin{abstract}
Single-spin asymmetries in the semi-inclusive production of charged pions in
deep-inelastic scattering from transversely and longitudinally polarized
proton targets are combined to evaluate the subleading-twist contribution to
the longitudinal case. This contribution is significantly positive for
\(\pi^+\) mesons and dominates the asymmetries on a longitudinally polarized
target previously measured by \hermes.  The subleading-twist contribution for
\(\pi^-\) mesons is found to be small. 
\end{abstract}

\pacs{13.60.-r, 13.88.+e, 14.20.Dh, 14.65.-q}

\maketitle

Single-spin asymmetries in the distribution of lepto-produced hadrons in the
azimuthal angle around the virtual photon direction are a valuable tool for
the exploration of transverse spin and momentum degrees of freedom in nucleon
structure. Whereas two out of the three fundamental quark distributions, the
unpolarized quark density and the helicity density, can be accessed in
inclusive measurements, this is not true for the remaining and so far
unmeasured transversity distribution
function~\cite{ph:RS,ph:ArtruM,ph:JJ}. Since transversity is chiral-odd and
since hard interactions conserve chirality, it can only be probed by a 
process involving some additional chiral-odd object. Single-spin asymmetries
in semi-inclusive deep-inelastic scattering (SIDIS), e.g., involving the
chiral-odd Collins fragmentation function~\cite{ph:Collins93}, could be the
required quark ``polarimeter'' to access transversity. This has been a main
motivation to look for azimuthal single-spin asymmetries. Such asymmetries
have been observed in SIDIS with unpolarized beams and with targets polarized
both longitudinally and transversely with respect to the beam
direction~\cite{hermes:azimp,hermes:azim0,hermes:azimd,hermes:h1prl1,
compass:a_ut_d}.
Asymmetries have also been observed with polarized beams and unpolarized
nucleons~\cite{Avakian:2003pk,Avetisyan:2004uz}. The asymmetry for a
transversely polarized target can be interpreted in terms of the transversity
distribution function, convoluted with the Collins fragmentation function, as
well as in terms of the Sivers~\cite{ph:Sivers} function, which appears with
the ordinary unpolarized fragmentation function. In the case of targets that
are polarized longitudinally with respect to the incoming beam direction, the
interpretation is more complex. In fact, the asymmetry for a target polarized
along the virtual photon direction contains various contributions from 
subleading-twist quark distribution and fragmentation functions. These 
contributions have -- for dynamics reasons -- an additional 1/\(Q\)
suppression compared to the ordinary 1/\(Q^4\) suppression of the Mott cross
section. When the polarization is along the beam direction, the small but
non-vanishing component of the nuclear spin transverse to the photon
direction, although 1/\(Q\) suppressed for kinematical reasons, also
contributes to the measured asymmetry. This feature has been exploited in
several
estimates~\cite{Korotkov:1999jx,Boglione:2000jk,Efremov:2001cz,Ma:2002ns,
Efremov:2003eq,Efremov:2003tf,Schweitzer:2003yr}
for the hitherto unknown transversity distribution  and Collins fragmentation
functions. However, some or all of the above-mentioned subleading-twist terms
have been neglected in all these estimates.

\begin{figure}
\includegraphics[width=5cm,angle=90]{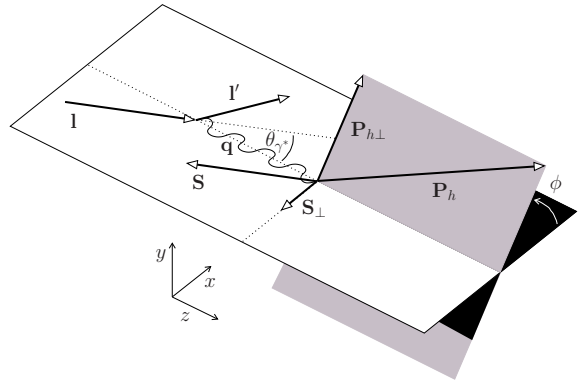}
\caption{\label{fig:phi} The definitions of the azimuthal angle \(\phi\) of
  the hadron production plane, relative to the plane containing the momentum
  \(\V{l}\) (\(\V{l'}\)) of the incident (scattered) lepton, the polar angle
  \(\theta_{\gamma^*}\) between the virtual photon and the incoming lepton
  directions, and of the transverse component \(\V{{S}_\perp}\) of the target
  spin \(\V{S}\) with respect to the photon direction \(\V{q}\equiv
  \V{l}-\V{l'}\).  }
\end{figure}

In this Letter the recently measured asymmetries on a transversely polarized
hydrogen target~\cite{hermes:h1prl1} are used to eliminate the contribution
due to the transverse spin component from the measured asymmetries on a
longitudinally polarized hydrogen target~\cite{hermes:azimp}, thereby
allowing, for the first time, the extraction of the purely subleading-twist
contribution. Knowledge of this subleading contribution is essential to any
extraction of information on the transversity distribution or Collins
fragmentation function from data with longitudinal target polarization.

Whenever the target is polarized with respect to the incoming beam direction
the measured asymmetries contain contributions from both the  transverse and
longitudinal polarization components with respect to  the virtual photon
direction. Throughout this paper asymmetries and their azimuthal moments will
carry one of the following superscripts for distinction: ``q'' when the
reference axis is the photon direction and ``l'' when it is the lepton
beam. They will be called photon-axis or lepton-axis asymmetries/moments,
respectively. Asymmetries and moments will also carry two-letter subscripts
denoting the polarization of beam and target. For the definition of azimuthal
angles, asymmetries, and azimuthal moments thereof the {\it Trento
  Conventions}~\cite{Bacchetta:2004jz} will be used.

The size of the component of the nucleon spin vector that is transverse to 
the virtual photon direction depends on  \(\theta_{\gamma^*}\), the polar
angle between the incoming beam direction and the virtual photon direction
(see Fig.~\ref{fig:phi}). Hence it strongly depends on the event kinematics.
At \hermes \ kinematics \(\sin\theta_{\gamma^*}\) can be as large as  15\%. 
In the configuration shown in Fig.~\ref{fig:phi}, where in the lab frame the
target spin vector is opposite to the incoming beam direction, the
transverse spin component lies in the lepton scattering plane. Therefore
\(\phi_S\), the azimuthal angle of the target spin relative to the scattering
plane, is equal to~\(\pi\).

The azimuthal moments of the distribution of hadrons around the 
virtual-photon direction can  be separated into contributions from the
longitudinal and transverse components of the target polarization with 
respect to the virtual photon direction. For the case of the longitudinal
lepton-axis moment \(\aulpmh\), the contributions from the transverse
component are the Sivers and Collins moments \(\smh\) and \(\cmh\). Since in
this case \(\phi_S=\pi\), both moments contribute to the $\sin\phi$ Fourier
component of the longitudinal lepton-axis asymmetry with a minus sign. In 
case of a transversely polarized target contributions arise from the
dominating transverse and from the small but non-vanishing longitudinal
component. Both the measured lepton-axis moments \(\cmph \) and \(\smph \)
contain contributions from \(\aulmh\). The lepton-axis moments are related 
to the photon-axis moments via the equation

\begin{widetext}
\begin{equation}\label{eq:matrix}
\left( \begin{array}{c}
\aulpmh\\[1.5mm]
\smph\\[1.5mm]
\cmph
\end{array} \right)
=
\left( \begin{array}{ccc}
\cos\theta_{\gamma^*}            & -\sin\theta_{\gamma^*}&-\sin\theta_{\gamma^*}  \\[1.5mm]
\frac{1}{2}\sin\theta_{\gamma^*} & \cos\theta_{\gamma^*} &      0        \\[1.5mm]
\frac{1}{2}\sin\theta_{\gamma^*} &       0               & \cos\theta_{\gamma^*} 
\end{array} \right)
\left( \begin{array}{c}
\aulmh\\[1.5mm]
\smh\\[1.5mm]
\cmh
\end{array} \right),
\end{equation}
\end{widetext}
which is valid up to corrections of order
\(\sin^2\theta_{\gamma^*}\)~\cite{diehl:2004}.

The complete analysis up to subleading-twist and leading order in $\alpha_s$
of longitudinal single-spin asymmetries in semi-inclusive DIS was presented 
in Ref.~\cite{Bacchetta:2004zf}, completing previous work of
Refs.~\cite{Mulders:1996dh,Boer:1998nt}. Neglecting quark mass effects, the
first term of the photon-axis moments on the right-hand-side of 
Eq.~\eqref{eq:matrix} is
\begin{widetext}
\begin{equation}
\aulmh = - \frac{(2-y)\sqrt{1-y}}{1-y+\frac{y^2}{2}}\, \frac{M}{Q}\,
\frac{
  \  \U{I}
  \! \left[
    \frac{\scriptVpperpu\cdot\scriptVkt}{M_h}\!\left(
\frac{M_h}{zM}g_{1L}{G^\perp} 
+ xh_L H_1^\perp 
      \right)\!
    +\!
    \frac{\scriptVpperpu\cdot\scriptVpt}{M}\!\left(
      \frac{M_h}{zM}h_{1L}^\perp {\tilde{H}} -xf_L^\perp D_1
      \right)
    \right]
}{f_1\ D_1}. \label{eq:subleading-twist} 
\end{equation}
\end{widetext}
The shorthand notation \(\U{I} \left[ \U{W} \, f \, D \right]\) is used here
for the convolution integral appearing in  the SIDIS cross section when quark
transverse momenta are included, i.e., 
\begin{eqnarray}
\U{I} \Big[ \U{W} \, f \, D \Big] \!\!\!&\equiv& \!\!\!
  \int \!\!\mathrm{d}^2 \V{P_{h\perp}} \, \mathrm{d}^{2} \V{p_{T}} \,
  \mathrm{d}^{2} \V{k_{T}} \ 
  \!\delta^{(2)}\!\!\left(\! 
               \V{p_{T}}- \frac{\V{P_{h\perp}}}{z} - \V{k_{T}}\!\! 
              \right) \nonumber \\
& & \times \Big[ \U{W} \, f(x, p_T^2) \ D(z, z^2k_T^2) \Big],
\end{eqnarray}
where \(\Vpperp\) is the transverse momentum of the detected hadron, \(\Vpt\)
(\(\Vkt\)) is the intrinsic quark transverse momentum in the generic
distribution function \(f\) (fragmentation function \(D\)), and \(\U{W}\) is 
a weight that depends on the involved distribution and fragmentation
functions. The masses \(m_q\), \(M\), and \(M_h\) are the quark, nucleon and
hadron masses and \(x\), \(y\), and \(z\) are the usual semi-inclusive DIS
Lorentz invariants. The quark charge squared weighted sum over the various
(anti)quark flavors and the dependence on \(x\) (\(z\)), \ \(p_{T}^2\)
(\(k_{T}^2\)), and \(Q^2\) of the distribution (fragmentation) functions have
been omitted in Eq.~\eqref{eq:subleading-twist}.

Since the extraction of the subleading-twist term \(\aulmh\) is the main
result of this work,  its components are  described  briefly. The asymmetry
arises from the interference of the scattering amplitudes of a longitudinal
and transverse photon. This leads to the specific dependence of the numerator
on the variable $y$. All terms in the numerator of
Eq.~\eqref{eq:subleading-twist} involve either combinations of
subleading-twist distribution functions ($h_L$, $f_L^\perp$) with
leading-twist fragmentation functions or of leading-twist distribution
functions in conjunction with subleading-twist fragmentation functions
($G^\perp$, $\tilde{H}$). One should note that it is not possible to give a
simple probabilistic interpretation to the subleading-twist functions.
The terms containing $h_L$ and $h_{1L}^\perp$ have been studied in some 
detail in Refs.~\cite{Boglione:2000jk,Efremov:2001cz}, making use of
Wandzura-Wilczek approximations, and of Lorentz covariance relations. 
(The latter have been proven to be not rigorous in Ref.~\cite{Goeke:2003az}.)
Note that the function $h_{1L}^\perp$ appears also in the $\sin 2 \phi$
Fourier component of the longitudinal single-spin
asymmetry~\cite{Mulders:1996dh}. Recent preliminary results of \clas \ 
support a non-vanishing $\sin 2 \phi$ moment~\cite{Avakian:2004qt}. However,
in measurements at \hermes \, which were in a different kinematic region than
the ones at \clas, it was found to be consistent with
zero~\cite{hermes:azimp}. The term with the helicity distribution $g_{1L}$
contains the fragmentation function \(G^\perp\), which is at present
unknown. A similar term appears also in the longitudinal beam-helicity
asymmetry, $A_{LU}$~\cite{Bacchetta:2004zf}. The latter has been found to be
non-zero both at \hermes~\cite{Avetisyan:2004uz} and at
\clas~\cite{Avakian:2003pk}. Finally, the last term of the numerator contains
the function $f_L^\perp$. Similar to the Sivers function \(f_{1T}^\perp\)
it is odd under time reversal (T-odd) and for this reason has been  neglected
in virtually all theoretical treatments of the measured lepton-axis
asymmetries on a longitudinally polarized target. Recently, it has been 
recognized that such T-odd distribution functions can arise through initial 
or final state interactions~\cite{ph:BHS,ph:Collins02,ph:JY,ph:BJY}. A
calculation of the function $f_L^\perp$ has been performed in
Ref.~\cite{Metz:2004je} in the context of a simple diquark spectator model. 
It should be noted that so far factorization has been proven only for
leading-twist observables in semi-inclusive deep-inelastic scattering with
hadrons in the current fragmentation region detected at low transverse
momentum~\cite{Ji:2004wu,Ji:2004xq}. A factorization proof for
subleading-twist observables is still open. At the moment no firm 
experimental information about any of the subleading-twist terms in
Eq.~\eqref{eq:subleading-twist} exist.

The extraction of \(\aulmh\) reported here gives a first indication about the
size of such subleading-twist effects in azimuthal target-spin
asymmetries. This is  especially important when measuring leading-twist
asymmetries of the same order of magnitude where the question arises whether
or not subleading-twist contributions can be neglected.

The other two terms on the right-hand-side of Eq.~\eqref{eq:matrix} read
\begin{eqnarray}
\smh \!&=& \! -\frac{
  \ \U{I}
  \! \left[
    \frac{\scriptVpperpu\cdot\scriptVpt}{2 M} f_{1T}^\perp D_1
    \right]
}{
f_1\,D_1}, \label{eq:Sivers}\\
\cmh \!&=& \! -\frac{1-y}{1-y+\frac{y^2}{2}}
  \frac{ 
  \ \U{I}
  \! \left[
    \frac{\scriptVpperpu\cdot\scriptVkt}{2 M_h} h_1 H_1^\perp
    \right]
}{ f_1\,D_1}\label{eq:Collins} 
\end{eqnarray}
which are the leading-twist Sivers and Collins moments, involving either the
Sivers distribution function, or the transversity distribution \(h_1\) in
conjunction with the Collins fragmentation function \(H_1^\perp\).

Measurements of azimuthal single-spin asymmetries on a transversely polarized
target can be used to eliminate the contribution of these leading-twist
moments~\eqref{eq:Sivers} and~\eqref{eq:Collins} to the longitudinal
lepton-axis moment \(\aulpmh\). Hence the subleading-twist terms
(Eq.~(\ref{eq:subleading-twist})) can be isolated. At \hermes\ kinematics the
deviation from unity of  \(\cos\theta_{\gamma^*}\) can be neglected. The
subleading-twist contribution \(\aulmh\) then reads 
\begin{eqnarray} 
\aulmh &=&  \aulpmh + \sin\theta_{\gamma^*} \cmph \nonumber \\
       & & + \sin\theta_{\gamma^*}\smph . \label{eq:hermes_t3-extraction}
\end{eqnarray}
Here \(\sin \theta_{\gamma^*}\) is evaluated from the lepton kinematics as 
\(2xMQ^{-1} \ \sqrt{1 - y - y^2x^2M^2/Q^2}/\sqrt{1 +  4x^2M^2/Q^2}\).

For the extraction of the subleading-twist contribution \(\aulmh\) according
to Eq.~\eqref{eq:hermes_t3-extraction} the lepton-axis asymmetries on a
longitudinally polarized hydrogen target~\cite{hermes:azimp} were reanalyzed
to have the same binning in \(x\) and \(z\) as in the measurement on a
transversely polarized hydrogen target~\cite{hermes:h1prl1}. Furthermore, the
\(\sin\phi\) modulation of the semi-inclusive cross section has been 
extracted in a fit of the normalized-yield asymmetry\footnote{This is in
  contrast to the previous publication on longitudinal single-spin
  asymmetries~\cite{hermes:azimp}  where a weighting method has been used to
  extract the \(\sin\phi\) Fourier component of the asymmetry.} 
\begin{equation}
A_{UL}^{\text{l}}(\phi)=\frac{1}{|P_L|}
\frac{N^\rightarrow(\phi)-N^\leftarrow(\phi)}{N^\rightarrow(\phi)+N^\leftarrow(\phi)},  
\end{equation}
where \(P_L\) is the longitudinal target polarization and
\(\rightarrow/\leftarrow\) denotes a target polarized antiparallel/parallel 
to the incoming beam direction. The same requirements on the lepton 
kinematics were used as in the analysis of the transverse target data, i.e.,
\(W^2 > 10\)\,GeV\(^2\), \(0.023 < x < 0.4\), \(0.1 < y < 0.85\) and 
\(Q^2>1\)\,GeV\(^2\), where \(W\) is the invariant mass of the initial
photon-nucleon system. Coincident hadrons were accepted only if \(0.2<z<0.7\)
and \(\theta_{\gamma^* h} > 0.02\)\,rad, where \(\theta_{\gamma^* h}\) is the
angle between the directions of the virtual photon and the hadron. The latter
requirement was imposed to avoid a region where the azimuthal angle \(\phi\)
is poorly reconstructed due to detector smearing. Pions were identified in 
the momentum range \(4\)~GeV \(<P_\pi<13.8\)~GeV using either a threshold
Cherenkov counter for the longitudinally polarized data set or a Ring Imaging
Cherenkov counter for the transversely polarized target data. The lepton-axis
moments from the transversely polarized target data set were extracted in the
fit  
\begin{eqnarray}
{A^{\text{l}}_{UT}(\phi,\phi_S)} 
    &=& \,2\, \cmph \ \sin(\phi + \phi_S) \nonumber\\
    & &\!\!\!\! + \, 2\, \smph \ \sin(\phi -\phi_S)\label{eq:A_UT-fit}
\end{eqnarray}
of the transverse asymmetry in Eq.~(1) of
Ref.~\cite{hermes:h1prl1}.\footnote{Note that in Ref.~\cite{hermes:h1prl1} a
  superscript on the asymmetry is used that is different and 
  not related to the one here. Also the fit in Ref.~\cite{hermes:h1prl1}
  includes kinematic prefactors that are not needed here.}

A possible uncertainty in the interpretation of the extracted
asymmetries in terms of Eq.~\eqref{eq:subleading-twist} is the contribution
to the analyzed pion samples  from the decay of exclusively produced vector
mesons (VM). Due to the limited acceptance of the \hermes \ spectrometer, a
large fraction of these vector mesons cannot be identified. Although the
contribution of their decay pions to the observed pion yield is small -- less
than 15\% for the highest \(z\) bin~\cite{hermes:h1prl1}, based on a 
\pythia6 \ Monte Carlo simulation tuned for \hermes \
kinematics reproducing the exclusive VM cross section on a 10\%
level~\cite{Liebing:2004us} -- their  
contribution to \(\smh \) for a transversely polarized target could be
significant~\cite{Goeke:2001tz}. For \(\aulmh\) this contributes only through
the transverse component and is thus subtracted through
Eq.~\eqref{eq:hermes_t3-extraction}. The VM contribution to the \(\aulmh \)
moments from the longitudinal spin component of the target can be treated as 
a dilution as no \(\sin\phi\) dependence  on the longitudinal target
polarization  of either the VM production or its decay distribution is
expected~\cite{Fraas:1974}. For an estimate of such effects moments were
extracted that have the diluting contribution from this exclusive channel
subtracted. This was done by dividing the \(\aulmh \) moments of
Eq.~\eqref{eq:hermes_t3-extraction} by \((1-N_{\mbox{\scriptsize
    VM}}/N_{\mbox{\scriptsize tot}})\) where \(N_{\mbox{\scriptsize VM}}\) 
and \(N_{\mbox{\scriptsize tot}}\) are the numbers of pions from VM decays 
and all detected pions, respectively.

\begin{table*}[ht]
\caption{The 2\(\aulmh\) moments of the \(\pi^+\) and \(\pi^-\) production
  cross section for different \(x\) (top) and \(z\) (bottom) bins. Only
  statistical uncertainties are included. In addition there is a common
  systematic uncertainty of 0.003. Results are shown for all detected pions
  and for the case where the contribution from the decay of exclusive VM has
  been subtracted.}\label{tab:a_ul-comparison}  
\begin{tabular*}{\textwidth}{cc|ccc|cc|cc}
\hline\hline
 & & & & &\multicolumn{2}{c|}{all \(\pi\)'s} & 
\multicolumn{2}{c}{\(\quad\pi\)'s from exclusive VM subtracted}\\
\(\quad \langle x\rangle \quad\) & \(\quad  \langle z\rangle \quad \) & 
\( \langle \pperp \rangle  \)[GeV] &\(\quad \langle y\rangle \quad\) & 
 \( \langle Q^2\rangle \)[GeV\(^2\)] &
\(\ \ \ 2\aulmpip \ \ \) & \(\ \ 2\aulmpim \ \ \) & 
\(\ \ \ \ 2\aulmpip \ \ \) & \(\ \ 2\aulmpim \ \ \) \\[1.3mm]
\hline
\( 0.038 \) & \( 0.36 \) & \( 0.50 \) & \( 0.68 \) & \( 1.3 \) & \(0.023\pm0.008\) & \(-0.012\pm0.010\) & \(0.025\pm0.009\) & \(-0.013\pm0.011\)\\ 
\( 0.067 \) & \( 0.41 \) & \( 0.45 \) & \( 0.59 \) & \( 2.0 \) & \(0.022\pm0.007\) & \(-0.012\pm0.010\) & \(0.023\pm0.008\) & \(-0.012\pm0.010\)\\ 
\( 0.114 \) & \( 0.43 \) & \( 0.42 \) & \( 0.55 \) & \( 3.2 \) & \(0.039\pm0.010\) & \(-0.016\pm0.013\) & \(0.041\pm0.010\) & \(-0.017\pm0.014\)\\ 
\( 0.178 \) & \( 0.44 \) & \( 0.41 \) & \( 0.52 \) & \( 4.8 \) & \(0.057\pm0.016\) & \(\ \ \,0.028\pm0.022\) & \(0.059\pm0.016\) & \(\ \ \,0.029\pm0.023\) \\ 
\( 0.274 \) & \( 0.46 \) & \( 0.40 \) & \( 0.48 \) & \( 6.8 \) & \(0.053\pm0.025\) & \(-0.023\pm0.035\) & \(0.054\pm0.025\) & \(-0.024\pm0.036\)\\
\hline
\( 0.065 \) & \( 0.26 \) & \( 0.42 \) & \( 0.71 \) & \( 2.3 \) & \(0.027\pm0.009\) & \(\ \ \,0.000\pm0.012\) & \(0.028\pm0.009\) & \(\ \ \,0.000\pm0.012\) \\ 
\( 0.080 \) & \( 0.35 \) & \( 0.45 \) & \( 0.62 \) & \( 2.5 \) & \(0.029\pm0.008\) & \(-0.018\pm0.011\) & \(0.030\pm0.009\) & \(-0.019\pm0.012\)\\ 
\( 0.091 \) & \( 0.47 \) & \( 0.48 \) & \( 0.55 \) & \( 2.4 \) & \(0.033\pm0.008\) & \(-0.002\pm0.011\) & \(0.035\pm0.009\) & \(-0.002\pm0.012\)\\ 
\( 0.098 \) & \( 0.62 \) & \( 0.49 \) & \( 0.49 \) & \( 2.3 \) & \(0.033\pm0.011\) & \(-0.021\pm0.015\) & \(0.037\pm0.012\) & \(-0.024\pm0.018\)\\ 
\hline\hline
\end{tabular*}
\end{table*}

The main contribution to the systematic uncertainty in the extracted moments 
arises from the measurement of the target polarization. Other contributions
include smearing due to detector resolution and radiative effects.
The combined systematic uncertainty is found to be less than 0.003.

\begin{figure}[t]
\includegraphics[width=0.48\textwidth,angle=0]{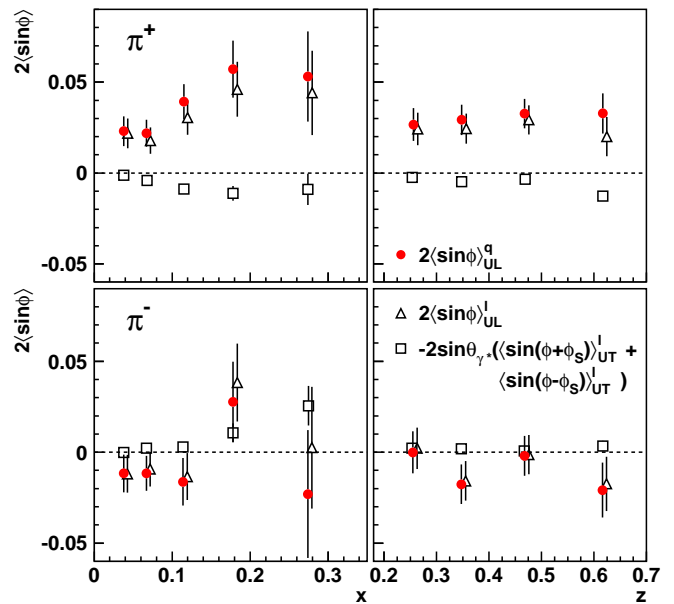}
\caption{\label{fig:beam-asym} The various azimuthal moments appearing in
 the measurement of the \(\sin\phi\) modulations of single-spin asymmetries 
 on a longitudinally polarized hydrogen target for charged pions as functions
 of \(x\) and \(z\). The open symbols are the measured lepton-axis
 moments. The ones from a transversely polarized target are multiplied by
 \(-\sin\theta_{\gamma^*}\) according to their appearance in the longitudinal
 lepton-axis moments. The closed symbol is the subleading-twist contribution
 to the measured lepton-axis asymmetries on a longitudinally  polarized
 target. The triangles are slightly shifted horizontally for distinction. An
 overall systematic error of 0.003 is not included in the figure.  
 }
\end{figure}

The moments  for charged pions are shown as functions of \(x\) and \(z\) in
Fig.~\ref{fig:beam-asym} and summarized in Table~\ref{tab:a_ul-comparison}. 
In addition to the extracted longitudinal photon-axis moments \(\aulmh\) 
the lepton-axis moments for longitudinally and transversely polarized targets
are plotted in Fig.~\ref{fig:beam-asym}. The latter include the prefactor
\(-\sin\theta_{\gamma^*}\)  with which they appear in the \(\aulpmh\)
measurement. The resulting longitudinal photon-axis moments are significantly
positive for the \(\pi^+\) and consistent with zero for the \(\pi^-\).  Hence
in the case of the \(\pi^+\) this subleading-twist contribution dominates the
measured lepton-axis asymmetries on a target that is polarized longitudinally
with respect to the beam direction. Therefore it becomes clear that those
asymmetries cannot be interpreted in terms of only the Collins fragmentation
function or the Sivers function. In particular, the contribution from the
Sivers function to the measured longitudinal lepton-axis asymmetries is small
compared to the subleading-twist contribution as it appears only for the
transverse component of the target spin. Unfortunately, due to the presence 
of several contributions (Eq.~\ref{eq:subleading-twist}), it is not possible
to make any statements about the size of any subleading-twist function
separately. Nevertheless, it is clear that subleading-twist effects cannot be
neglected a priori. This will be important when interpreting the measured
lepton-axis asymmetries on a transversely polarized target which for
experimental reasons receive not only contributions from the transverse 
target spin component (e.g., the Collins and Sivers effects) but also from 
the longitudinal component (subleading-twist) as in
Eq.~\eqref{eq:matrix}.\footnote{It should be noted that a similar analysis 
  for the azimuthal moments on a transversely polarized target yields 
  corrections to the measured lepton-axis moments~\cite{hermes:h1prl1} that
  are negligible compared to their statistical uncertainty.}

In summary, single-spin asymmetries on hydrogen polarized longitudinally 
along the photon direction have been extracted for the first time. 
The contribution to the lepton-axis asymmetries from the transverse spin
component in the measurement on a target polarized longitudinally with 
respect to the beam has been subtracted using the data from a transversely
polarized hydrogen target. The averaged asymmetries in the range
\(0.023<x<0.4\) (\(\langle x\rangle = 0.082\)) and \(0.2<z<0.7\) 
(\(\langle z \rangle =0.40\)) are \(0.030\pm 0.004_{\mbox{\scriptsize ~stat}}
\pm 0.002_{\mbox{\scriptsize ~sys}}\) for \(\pi^+\) and \(-0.009 \pm 
0.006_{\mbox{\scriptsize ~stat}} \pm 0.001_{\mbox{\scriptsize ~sys}}\) for
\(\pi^-\). For \(\pi^+\) the \(\aulmh\) result is the dominating component in
this range. This shows that subleading-twist effects are large and can, at
\hermes\ kinematics, be comparable to leading-twist effects. This must be
taken into account when interpreting asymmetries on transversely or 
longitudinally polarized targets solely in terms of leading-twist
functions. At lower energies the isolation of leading-twist effects may be
even more difficult.

\begin{acknowledgments} 
We thank M.~Diehl and F.~Pijlman for many interesting discussions. 
We gratefully acknowledge the \desy\ management for its support and the staff
at \desy\ and the collaborating institutions for their significant
effort. This work was supported by  
the FWO-Flanders, Belgium; 
the Natural Sciences and Engineering Research Council of Canada;
the National Natural Science Foundation of China;
the Alexander von Humboldt Stiftung;
the German Bundesministerium f\"ur Bildung und Forschung;
the Deutsche Forschungsgemeinschaft (DFG);
the Italian Istituto Nazionale di Fisica Nucleare (INFN);
the Monbusho International Scientific Research Program, JSPS,
and Toray Science Foundation of Japan;
the Dutch Foundation for Fundamenteel Onderzoek der Materie (FOM);
the U.~K.~Engineering and Physical Sciences Research Council and the
Particle Physics and Astronomy Research Council;
and the U.~S.~Department of Energy and the National Science Foundation.
\end{acknowledgments}

\bibliography{twist-3}

\end{document}